\documentclass[11pt]{article}
\usepackage[normalem]{ulem}
\pdfoutput=1 % if your are submitting a pdflatex (i.e. if you have

\usepackage{jheppub} % for details on the use of the package, please
\usepackage{url}
\usepackage{caption}
\usepackage{subcaption}
\usepackage{extarrows}

\usepackage[latin9]{inputenc}
\usepackage{float}
\usepackage{amsmath}
\usepackage{amssymb}
\usepackage{graphicx}
\usepackage{esint}
\usepackage{hyperref}
\usepackage{comment}
\usepackage{color}
\usepackage{microtype}
\usepackage{cleveref}
\usepackage{breakurl}
\usepackage{bbm}
\newcommand{\be}{\begin{equation}}
\newcommand{\ee}{\end{equation}}
\newcommand{\ben}{\begin{displaymath}}
\newcommand{\een}{\end{displaymath}}
\newcommand{\bea}{\begin{eqnarray}}
\newcommand{\eea}{\end{eqnarray}}
\def\K{K{\"a}hler }
   \newcommand{\rf}[1]{(\ref{#1})}
\newcommand{\vp}{\varphi}

\def\be{\begin{equation}}
\def\ee{\end{equation}}
\def\bea{\begin{eqnarray}}
\def\eea{\end{eqnarray}}
\def\ba{\begin{array}}
\def\ea{\end{array}}
\def\bit{\begin{itemize}}
\def\eit{\end{itemize}}

\def\a{\alpha}

\def\vp{\varphi}

\def\rmi{{\rm i}}

\newcommand{\cN}{\mathcal{N}}

\newcommand{\ft}[2]{{\textstyle\frac{#1}{#2}}}
\def\rmi{{\rm i}}

\def\rme{{\rm e}}

 \makeatletter

\allowdisplaybreaks

\makeatother

\makeatletter
\DeclareRobustCommand{\rcite}[1]{%
  \rcite@aux#1,\@nil{#1}%
}
\def\rcite@aux#1,#2\@nil#3{%
  \if\relax#2\relax
    Ref.~\cite{#3}%
  \else
    Refs.~\cite{#3}%
  \fi
}
\makeatother

\hypersetup{
    colorlinks = true,
    citecolor = {blue},
    linkcolor = {blue},
    urlcolor = {blue},
}

\def\eqn#1{eq.~\eqref{#1}}

\def\rcite#1{ref.~\cite{#1}}

 \title{\rm {  \bf   Jordan Frame in Supergravity  and  Cosmology
}
}

\author{Renata Kallosh}

\affiliation{Leinweber Institute for Theoretical Physics at Stanford, 382 Via Pueblo, Stanford, CA 94305, USA}
\emailAdd{kallosh@stanford.edu}
\abstract{   
The superconformal action can be gauge-fixed in a gauge where is leads to the Einstein frame supergravity defined by a  \K potential $\mathcal{K}(z, \bar z)$, or in a gauge where it leads to a Jordan frame supergravity defined by the frame function $\Omega(z, \bar z)$, in addition to  $\mathcal{K}(z, \bar z)$.  We present {\it new supergravity $\xi$-attractor models with non-minimal gravity coupling}. They describe potentials with an exponential and a polynomial approach to the plateau.  The  previously known $\xi$-attractors in the large $\xi$ limit
predicted the spectral index $n_{s} = 1-2/N$ and the tensor-to-scalar ratio $r\to {12\over N^2}$. The new $\xi$-attractors  in the large $\xi$ limit 
predict $n_{s} \geq 1-2/N$ and $r\to 0$, which provides a better match to the recent observational data and makes these models a target for the future B-mode experiments with small $r$.

New $\xi$-attractors have some features similar to those of Palatini attractors. However, we show that the Palatini gravity with nonminimal scalar coupling and an independent affine connection has no supergravity embedding.  }

\begin{document}

\maketitle

% \tableofcontents{}
 
\parskip 7 pt

%\newpage
\section{Introduction} 
The non-minimal coupling of scalars to gravity in the context of cosmology has been studied for a long time, since it was discovered in \cite{Salopek:1988qh,Bezrukov:2007ep} that the Higgs boson of the Standard Model can lead to inflation if the Higgs scalar has a non-minimal coupling to gravity. The supersymmetric versions of the standard model, MSSM and NMSSM, were also constructed. These are models with a global supersymmetry. But in applying particle physics to cosmology, one must include gravity; therefore, a supersymmetric version of particle physics for cosmology had to be formulated within the context of supergravity. The supergravity models with non-minimal coupling of Higgs bosons to gravity were constructed as a specific example of the general type Jordan frame supergravities \cite{Ferrara:2010yw,Ferrara:2010in}.

This construction of $\cN=1$ Jordan frame supergravities is based on superconformal theory \cite{Kallosh:2000ve}.  There is a gauge-fixing of the Weyl symmetry in which supergravity is in the Einstein frame, and the scalars are minimally coupled to gravity. There is another gauge in which supergravity is in the Jordan frame, and the scalars are non-minimally coupled to gravity.

The superconformal origin of  Poincar\'e supergravity theory was discovered long time ago \cite{Kaku:1977pa,Ferrara:1977ij,Ferrara:1978jt,Ferrara:1978wj,Townsend:1979ki}. A consistent extended conformal supergravity was developed in \cite{Bergshoeff:1980sw,Bergshoeff:1980is,VanProeyen:1983wk}. A general $\cN=1$ supergravity with matter multiplets was derived in \cite{Cremmer:1982en} based on a superconformal calculus. Relevant to our discussion was a particular feature of this construction. The frame function $\Omega$  defines the \K potential in a unique way, namely
\be
\Omega(z, \bar z) = e^{-{1\over 3} \mathcal{K}(z, \bar z)} \ .
\label{frameK}\ee
About two decades later, motivated by theoretical cosmology,  a general $SU(2,2|1)$ superconformal theory was presented in \cite{Kallosh:2000ve} and the procedure of the gauge-fixing extra gauge symmetries was proposed to produce a general Poincar\'e supergravity.  The main new feature of the construction in \cite{Kallosh:2000ve} was the fact that the relation between  the frame function and the \K potential was not  unique anymore, and in fact, these were two independent functions
\be
\Omega(z, \bar z) =y\, \bar y \, e^{-{1\over 3} \mathcal{K}(z, \bar z)} \ .
\label{frameKy}\ee
The set up in \cite{Kallosh:2000ve} involved  $(n+1)$   complex coordinates  $X^I, \bar X^{\bar I}$  in superconformal theory with a  \K metric with the signature $- + \dots +$. The negative direction corresponds to a scalar compensator. The holomorphic coordinates $X^I$ are defined by a set $y, z^\a$ where the $n$ coordinates $z^\a$ of the projective \K manifold with a \K metric with the signature $ + \dots +$  are the scalars of a Poincar\'e supergravity,
\be
X^I = y \, Z^I(z) \ .
\ee
 For example, in a gauge $Z^0=1, Z^\a= z^\a$, the meaning of the modulus $y$ is the gauge-fixed value of the conformon field  $X^0=y$. As a result, the superconformal theory admits a gauge-fixing with a Jordan frame supergravity \cite{Ferrara:2010yw,Ferrara:2010in} where there are {\it two independent  real functions of} $(z, \bar z)$
\be
\Omega(z, \bar z) \, , \qquad  \mathcal{K}(z, \bar z) \ .
\label{two}\ee 
not satisfying a constraint in eq. \rf{frameK}, but the one in eq.  \rf{frameKy}. Note that here $\Omega(z, \bar z)$ is a frame function defining a choice of the  Jordan frame, whereas the \K potential $\mathcal{K}(z, \bar z)$ is a \K potential of the Einstein frame. In fact, there is no \K potential in the Jordan frame, and the kinetic term of scalars is not defined by a \K metric.

The explicit general theory of supergravity in an arbitrary Jordan frame was derived in \cite{Ferrara:2010yw, Ferrara:2010in} by a gauge fixing of the $SU(2,2|1)$ superconformal theory.    This was motivated by non-minimal scalar-gravity couplings and by supergravity-based cosmological models, including supersymmetric Higgs inflation. 

 A detailed, clear derivation of supergravity from superconformal theory, based on improved superconformal calculus, is presented in the textbook \cite{Freedman:2012zz}, with additional references. We describe in Sec. \ref{Sec:con} the conditions for a consistent superconformal theory and in Sec. \ref{Sec:JE} the Jordan and Einstein frames of supergravity as obtained from two different gauges of the superconformal theory.

In Secs. \ref{Sec:cosmology} and \ref{Sec:1sc}, we describe the known cosmological applications of supergravity in Jordan and Einstein frames.
Some of the cosmological models with the Jordan frame function $\Omega=1+\xi \phi^n$, known as  $\xi$-attractors, were proposed in   \cite{Kallosh:2013tua,Galante:2014ifa}. They were related to cosmological exponential $\a$-attractors \cite{Ferrara:2013rsa,Kallosh:2013yoa} which were in good agreement with cosmological data.

Recently, the cosmological data changed \cite{Balkenhol:2025wms} and attracted attention to cosmological polynomial  $\a$-attractors \cite{Kallosh:2022feu}, which so far were not known in the form of the $\xi$-attractors in models with non-minimal coupling of scalars to gravity. The review of the recent status of inflationary models is presented in \cite{Kallosh:2025ijd}. Both polynomial and exponential attractor models are interesting due to the remaining uncertainty in the combination of various cosmological data sets \cite{Ferreira:2025lrd}.
 
Our main goal in this paper is to find a set of new $\xi$-attractors in models with non-minimal coupling of scalars to gravity. These should have a supergravity embedding and include both exponential and polynomial attractor models, which are counterparts of the corresponding exponential and polynomial attractors in models with minimal coupling to gravity, reviewed in \cite{Kallosh:2025ijd}. A crucial role will be played by the fact that the Jordan frame supergravity \cite{Ferrara:2010yw,Ferrara:2010in} is defined by two independent functions of physical scalars, as shown in eq. \rf{two}. The supergravity version of the new $\xi$-attractors is presented in Secs. \ref{Sec:new} and \ref{Sec:pol}. The known supergravity cosmological models of early $\xi$-attractors in  \cite{Kallosh:2013tua} are given in Appendix \ref{A:Hxi} for convenience. 

 In \cite{Ferrara:2010yw,Ferrara:2010in}, the Jordan/Einstein frame supergravity depends on the Jordan/Einstein metric, and the affine connections are metric dependent. The last condition is not satisfied, e.g. in the Palatini versions of the scalar fields coupled to gravity, where the affine connections are metric-independent, see \cite{Tenkanen:2020dge} for a review and extensive bibliography. 
We therefore study the most general internally consistent Jordan-frame supergravities in  Sec. \ref{Sec:con} and apply it to the most recent work \cite{Pallis:2026cyz}. In Appendix \ref{A:Pal} we  argue that the cosmological Palatini attractors with a metric-independent affine connection cannot be promoted to a consistent supergravity level involving fermions.  An updated version of \cite{Pallis:2026cyz} agrees with our conclusion that Palatini gravity has no supergravity embedding. We also agree that bosonic Palatini cosmological models were useful in clarifying the status of non-minimal-coupling supergravity models.

In  \cite{Kallosh:2013tua,Galante:2014ifa} the choice of $\xi$-attractor cosmological models started with a choice of the Jordan frame function $K_J(\phi)$ defining the scalar kinetic term ${1\over 2} K_J(\phi) (\partial \phi)^2$. Our new choice of cosmological models starts with the choice of the  \K potential $\mathcal{K}(z, \bar z)$ and \K metric; for a single complex scalar the kinetic terms is $g_{z \bar z}\partial z \partial \bar z= \frac{\partial^2  \mathcal{K} (z, \bar z)}{\partial z  \partial \bar z} \partial z \partial \bar z$. Therefore, the choice of the new $\xi$-attractor cosmological models starts with a choice of the Einstein scalar kinetic term ${1\over 2} K_E(\phi) (\partial \phi)^2$ where $K_E(\phi)={1\over 2} g_{z \bar z}|_{z=\phi}$. 
We find that these models have significant advantages over the old $\xi$-attractor models.

\noindent 1. The ones with  scalars coupled to gravity as $\xi\phi^2$, and a new choice we made for $K_E, K_J, V_J$, have a simple relation to exponential and polynomial $\a$-attractors  with hyperbolic geometry \cite{Kallosh:2013yoa,Kallosh:2022feu} iff
\be
\xi={1\over 6\a} \ .
\label{xia}\ee
2. The polynomial $\xi$-attractors with scalars coupled to gravity as $\xi\phi^n$, and a  choice we made for $K_E, K_J, V_J$,  are related to general KKLTI type attractors  \cite{Martin:2013tda,Kallosh:2018zsi,Kallosh:2019hzo} 
 with parameters $(k, \mu)$ iff
\be
 n={2(k+1)\over k} >2, \ , \quad n={2k\over 2+k} < 2, \, \quad \xi= {k^2\over \mu^2}  \ .
 \label{xin}\ee
The relation between known exponential and polynomial $\a$-attractors and general KKLTI type attractors to new $\xi$-attractors (with choices we made for $K_E, K_J, V_J$) is simple: they are different gauges of the superconformal theory under conditions that \rf{xia} and \rf{xin}, respectively,  are satisfied.

\section{Consistency of a superconformal theory}\label{Sec:con}

In the case of $\cN=1$ superconformal algebra $SU(2,2|1)$, for each generator of the algebra 
\be
P_a, \, \, M_{ab}, \, \, D, \, \, K_a, \, \, T, \, \, Q, \, \, S\ee
 one needs a parameter and a gauge field. For example, for translation $P_a$, the parameter is $\xi^a$, and the gauge field is a vierbein $e_\mu^a$. For the Lorentz generator 
$M_{ab}$ the parameter is $\lambda^{ab} $ and the gauge field is a spin connection $\omega _\mu^{ab}$, for $Q$ supersymmetry the parameter is $\epsilon$ and the gauge field is gravitino $\psi_\mu$,  see Table 16.1 in \cite{Freedman:2012zz} for all generators and parameters and gauge fields. The whole set of gauge fields includes
\be
\,  \, e_\mu^a, \, \, \omega _\mu^{ab}, \, \, b_\mu, \, \, f_\mu^a, \, \, A_\mu, \,  \psi_\mu, \, \, \phi_\mu 
\label{gauge} \ee
There are  problems with this set of independent gauge fields which were recognized very early  \cite{Kaku:1977pa,Ferrara:1977ij,Ferrara:1978jt,Ferrara:1978wj,Townsend:1979ki,Bergshoeff:1980sw,Bergshoeff:1980is,VanProeyen:1983wk}. It was realized  that 
Weyl's theory of gravity may be regarded as the gauge theory of the conformal group. However, there is a subtlety: there is a translation operator $P_a$ in the group, but in gravitational theories it has to be replaced by a {\it covariant} general coordinate transformation; see \cite{VanProeyen:1983wk,Freedman:2012zz} for details. 

Specifically, the gauge symmetry transformations form an algebra, where two local (space-time dependent) supersymmetries  produce a  covariant general coordinate transformation (cgct) plus other terms
\be
[\delta_Q(\epsilon_1), \delta_Q(\epsilon_2)] = \delta_{cgct} \left (\xi^\mu= {1\over 2} \bar \epsilon_2 \gamma^\mu \epsilon_1\right)+\dots
\ee
This is different from the superconformal group, where two supersymmetry generators produce a  translation operator
\be
[\delta_Q(\epsilon_1), \delta_Q(\epsilon_2)] = -{1\over 2} \bar \epsilon_2 \gamma^a \epsilon_1 P_a \ .
\ee
Moreover,  in the supersymmetric context, it was found that the number of the off-shell bosonic degrees of freedom is not equal to the number of  fermionic degrees of freedom in the case that all fields in  \rf{gauge} are independent, for example, if $\omega _\mu^{ab}$ is independent of the vierbein $e_\mu^a$,
\be
N_{bos} -N_{ferm} \neq 0 \, .
\label{bf}\ee
The solution to these problems was presented in 
\cite{Kaku:1977pa,Ferrara:1977ij,Ferrara:1978jt,Ferrara:1978wj,Townsend:1979ki,Bergshoeff:1980sw,Bergshoeff:1980is,VanProeyen:1983wk}  in the form of the constraints on 3 types of curvatures,  in  $P_a, M_{ab}, Q$ directions
\bea
&&R_{\mu\nu}(P^a)= 0 \ , \cr 
&&e^\nu_b \hat R_{\mu\nu}(M^{ab})= 0 \ , \cr
&&\gamma^\mu  R_{\mu\nu}(Q)= 0 \ .
\label{curv}\eea 
This is eq. (16.12) in \cite{Freedman:2012zz}, and all detailed notations explaining these constraints are in the book. When these constraints are imposed on the gauge fields in \rf{gauge}, one finds that some gauge fields in \rf{gauge} are not independent anymore.
The independent fields are
\be
e_\mu^a, \, \, b_\mu,  \, \, A_\mu, \, \, \psi_\mu  \ .
\label{gaugeC} \ee
In particular, one finds from the constraint $R_{\mu\nu}(P^a)= 0 $ that the spin connection is now a function of the independent gauge fields in eq. \rf{gaugeC}
\be
\omega _\mu^{ab}|_{\rf{curv}}= \omega _\mu^{ab} (e, b, \psi)= \omega _\mu^{ab} (e)+ \omega _\mu^{ab} ( b)+ \omega _\mu^{ab} (\psi) \ .
\label{SC}\ee
Here
\be
\omega _\mu {}^{ab}(e)= 2 e^{\nu[a} \partial_{[\mu} e_{\nu]}{}^{b]} -
e^{\nu[a}e^{b]\sigma} e_{\mu c} \partial_\nu e_\sigma{}^c\,.
\label{omegae}\ee

When the constraints \rf{curv} are imposed, {\it there is an equal number of boson and fermion
fields, and the gauge symmetry algebra, including covariant general coordinate transformation, is consistent}.

The remaining gauge fields include bosonic fields, $b_\mu$ and $A_\mu$, in addition to a gravitational field $e_\mu^a$ and a gravitino $\psi_\mu$. Using the special conformal transformation $K_a$ one can fix 
$
b_\mu=0 \ .
$.
One also finds that, on-shell, the vector field $A_\mu$ depends on matter fields; see eq. \rf{A} below. This leaves us at the linear level with just two gauge fields for the on-shell linearized Weyl supermultiplet
\be
e_\mu^a, \, \,  \psi_\mu \ .
\label{gaugeCl} \ee
In this way, a perfect agreement is reached with the earlier Ferrara-Zumino construction of the superfield formulation of linearized conformal supergravity \cite{Ferrara:1977mv}. Note that in superspace,\footnote{It was noticed in \cite{Townsend:1979ki} that the second order formalism arises naturally in the superspace and therefore $\omega _\mu^{ab}= \omega _\mu^{ab}(e)$ is consistent with boson-fermion balance.} the fermion-boson balance is automatic, and the on-shell chiral linearized Weyl multiplet in the chiral basis is \cite{Ferrara:1977mv, Kallosh:1980fi,Freedman:2017zgq}
\be
W_{\alpha \beta\gamma}(y,\theta)= \psi_{\alpha \beta\gamma} +\theta^\delta  C_{\alpha \beta\gamma\delta} \ .
\ee 
Here $ \psi_{\alpha \beta\gamma}$ is the totally symmetric spinor gravitino field strength and $C_{\alpha \beta\gamma\delta}
$ is the totally symmetric spinor, the spinorial equivalent of the Weyl
conformal tensor.

Thus, we have learned that the Weyl supermultiplet is consistent only if the spin connection depends on the vierbein, as shown in eq. \rf{omegae}.

\section{Supergravity in the Jordan and  Einstein  frames}\label{Sec:JE}

In the previous section, we have confirmed that {\it  supergravity models in the Jordan frame depend on the Jordan metric; the affine connections are metric-dependent}. This is a feature of the Jordan frame in supergravity \cite{Ferrara:2010yw,Ferrara:2010in}, no other constructions are available, and they are not even possible without violating the simple supersymmetry rule about the balance between bosonic and fermionic degrees of freedom.

We will bring up the properties of Jordan frame supergravity here, which clarify its relation to cosmological attractor models.
The general theory of supergravity in an arbitrary Jordan frame was derived in \cite{Ferrara:2010yw,Ferrara:2010in} by a gauge fixing of the $SU(2,2|1)$ superconformal theory.   

The extra gauge symmetries of the superconformal theory, not present in the Poincar\'e supergravity, include a Weyl symmetry, which is a 
local conformal symmetry, which rescales the metric and allows us to derive the supergravity action either in the Einstein
frame or in an arbitrary Jordan frame.

Note that in both of these frames, the Poincar\'e supergravity derived from superconformal theory is in a second-order formalism where  the space-time curvature $R$ is defined by the torsionless spin connection 
 $\omega _\mu^{ab}(e)$ in a gauge where the special conformal transformations are fixed with $b_\mu=0$.

The Einstein frame Lagrangian,
in units of $M_P=1$,   is ${\cal L}_E= e_E  { 1\over 2} \,R
((e^a_\mu)_E)+\dots$, there is no direct scalar-curvature coupling. The Jordan
frame Lagrangian is ${\cal L}_J= e_J\,  { \Omega(z, \bar z)\over
2} \, R ((e^a_\mu)_J)+\dots$, where $\Omega(z, \bar z)$ is an arbitrary function of
complex scalar fields  $z, \bar z$. Therefore, in general, there is a
scalar-curvature coupling in the Jordan frame.  The vierbeins in these two frames are related by a Weyl transformation 
\be
(e^a_\mu)_J = \Omega(z, \bar z)^{1/2} (e^a_\mu)_E \ .
\ee
A local superconformal symmetry allows us to make a choice of a frame function, an arbitrary  Kahler potential, and a potential
\be
\Omega (z, \bar z)=1, \qquad \mathcal{K}(z, \bar z),   \qquad V_E(z, \bar z)\ , \ee
 to get the
Einstein frame supergravity. Here
\begin{equation}
V_{E}=V_{E}^F +V_{E}^D = \rme^{\mathcal{K}}\left( -3W\overline{W}
+\nabla _\alpha Wg^{\alpha \bar\beta }\nabla _{\bar\beta }\overline{W}
\right) +  \ft12( {\rm Re}
f)^{-1\,AB}P_AP_B\,, \label{VE}
\end{equation}
see eq. (18.11) in the textbook \cite{Freedman:2012zz} explaining all notations in $V_{E}$ here.
When one of the chiral superfields is a nilpotent one, the corresponding supergravity theory is described in \cite{Kallosh:2025dac} and references therein. The potential for the chiral multiplets in the presence of the nilpotent multiplet is not defined by Eq. \rf{VE} anymore and can be given by an arbitrary function $V_{E}(z, \bar z)$. It is also a property of liberated supergravity \cite{Farakos:2018sgq} that the scalar potential is augmented by the arbitrary function, which allows freedom in constructing phenomenological models rooted in supergravity.

In gauges with an arbitrary   frame function and with a  specific Jordan frame moduli space metric $(g_{_J})_{\alpha \bar \beta}$ and a potential $V_J$
\be
\Omega = \Omega(z, \bar z)\, , \quad (g_{_J})_{\alpha \bar \beta} = \Omega \, \mathcal{K}_{\alpha \bar \beta} -3 \frac{\Omega
_\alpha \Omega _{\bar\beta }}{\Omega }\, ,   \qquad V_J(z, \bar z)= \Omega^2 V_E  \ ,
\label{rule1}\ee
we get the Jordan frame supergravity.

The scalar-gravity part\footnote{The total action including vectors and fermions is given in Sec. 2 of   \cite{Ferrara:2010yw}.}  of the ${\cal N}=1$, $d=4$ supergravity in a generic Jordan frame depends on a frame function $\Omega(z, \bar z)$, on a  \K\, potential $\mathcal{K}(z, \bar z)$ independent on the frame function, and on a potential $V_J(z, \bar z)= \Omega^2 V_E$ where, in case with a nilpotent multiplet, the potential  $V_E(z, \bar z)$ is arbitrary. The scalar-gravity part of the action  is, according to \cite{Ferrara:2010yw},
\begin{eqnarray}
\mathcal{L}_{J}^{\rm sc-gr} =e_J\left[\Omega \left(  \frac{1}{2} {R}\left ((e^a_\mu)_J\right)-3 {\cal A}^2_\mu(z, \bar z) \right )
-\left( \Omega g_{\alpha\bar\beta }-3\frac{\Omega
_\alpha \Omega _{\bar\beta }}{\Omega }\right)  {\partial}_\mu z^\alpha  {\partial}^\mu \bar z^{\bar\beta }
-V_J \right]\, .
  \label{Turin-4}
\end{eqnarray}
In the absence of fermions, we can use the metric, $g_{\mu\nu}= e^a_\mu \eta_{ab} e^b_\nu$ instead of the vierbeins, and again, we remind that the scalar space-time curvature $R$ depends only on the metric in both frames. Here
\begin{eqnarray}
&& \Omega_\alpha \equiv {\frac{\partial }{\partial z^\alpha }}\Omega(z, \bar z) \,, \qquad  \Omega_{\bar\beta } \equiv {\partial\over \partial{\bar z}^{\bar\beta }}  \Omega(z, \bar z)=\overline{\Omega} _{\bar \beta} \, ,
\nonumber\\
&& g_{\alpha \bar \beta}= {\frac{\partial^2  \mathcal{K}_E(z, \bar z)}{\partial z^\alpha  \partial \bar z^{\bar \beta}}}\equiv \mathcal{K}_{\alpha \bar \beta} (z, \bar z)\, ,
\label{Km}\end{eqnarray}
and $\mathcal{A}_\mu$ is the purely bosonic part of the on-shell value of the auxiliary field $A_\mu $. On shell and in the absence of the vector fields, it depends on the scalar
fields as follows:
\be {\cal A}_\mu(z,{\bar z}) \equiv
-\frac{\rmi}{2\Omega }\,
  \left(  {\partial}_\mu z^\alpha\partial_\alpha\Omega
  -  {\partial} _\mu \bar z^{\bar\alpha }\partial_{\bar\alpha }\Omega \right).
 \label{A} \ee
The case $\Omega=1, \, {\cal A}_\mu=0\, ,(e^a_\mu)_J=(e^a_\mu)_E$ in eq. \rf{Turin-4} provides an action in the Einstein frame
\begin{eqnarray}
\mathcal{L}_{E}^{\rm scalar-grav} = e_{E}\left[   \frac{1}{2} {R}((e^a_\mu)_E)
-\mathcal{K}_{\alpha \bar \beta} (z, \bar z)  {\partial}_\mu z^\alpha  {\partial}^\mu \bar z^{\bar\beta }
-V_E \right]\, ,
  \label{Stanford}
\end{eqnarray}
and we require 
\begin{equation}
g_{\a \bar \beta}= \mathcal{K}_{\alpha \bar \beta} (z, \bar z)\, 
>0\,.
\label{Kibark}
\end{equation}

\section{Cosmology applications}\label{Sec:cosmology}

Consider the case of one physical scalar field $(z, \bar z)$ and one nilpotent superfield $S$ with $S^2=0$. The role of the nilpotent superfield is to allow an arbitrary choice of the potential $V(z, \bar z)$, see for example \cite{Kallosh:2025dac}. Therefore, in this discussion, we will consider only one physical scalar superfield $(z, \bar z)$ with \K potential $\mathcal{K}$ and make an arbitrary choice of the potential $V_E(z, \bar z)$ instead of the one in eq. \rf{VE}.

From superconformal theory, we learned that in gauges with an arbitrary   frame function $\Omega$ and with a  specific choice of the \K potential $\mathcal{K} $ and a potential $V_J$
\be
\Omega = \Omega(z, \bar z)\, , \quad \mathcal{K}= \mathcal{K} (z, \bar z)\, ,   \quad V_J(z, \bar z)= \Omega^2 V_E (z, \bar z) \ ,
\label{rule}\ee
we get the Jordan frame supergravity  \cite{Ferrara:2010yw}
\begin{eqnarray}
\mathcal{L}_{J}=\sqrt {-g_J}\left[\Omega \left(  \frac{1}{2} {R}\left (g_J\right)-3 {\cal A}^2_\mu(z, \bar z) \right )
-(g_{_J})_{z \bar z}  {\partial}_\mu z  {\partial}^\mu \bar z
-V_J (z, \bar z) \right]\, ,
  \label{Turin-41}
\end{eqnarray}
where\footnote{Note that the term $-3 \frac{\Omega
_z \Omega _{\bar z }}{\Omega }$ in the Jordan frame moduli space metric $(g_{_J})_{z \bar z}$ is required since under the Weyl transformation of the metric, the curvature $R(g_J)$ transforms.
In Palatini formalism the Ricchi curvature $ R_{\mu\nu}^{P} ( \Gamma)$ is Weyl invariant and therefore $(g_{_J})_{z \bar z}^{P} = \Omega \, \mathcal{K}_{z \bar z}$.}
\be 
(g_{_J})_{z \bar z} = \Omega \, \mathcal{K}_{z \bar z} -3 \frac{\Omega
_z \Omega _{\bar z }}{\Omega }\, ,\qquad 
{\cal A}_\mu(z,{\bar z}) \equiv
-\frac{\rmi}{2\Omega }\,
  \left(  {\partial}_\mu z\partial_z \Omega
  -  {\partial} _\mu \bar z\partial_{\bar z}\Omega \right) \ ,
 \label{A1} \ee
 and 
\be
\mathcal{K}_{z \bar z} (z, \bar z) \equiv g_{z \bar z}= {\frac{\partial^2  \mathcal{K}(z, \bar z)}{\partial z  \partial \bar z}}>0\, , \quad \Omega_z \equiv {\frac{\partial }{\partial z }}\Omega(z, \bar z) \,, \quad  \Omega_{\bar z } \equiv {\partial\over \partial{\bar z}^{\bar z }}  \Omega(z, \bar z) \ .
\ee 
Therefore, for cosmological applications in supergravity, it is natural to start with the choice of the \K potential $ \mathcal{K}(z, \bar z)$ and \K metric $g_{z\bar z}= \mathcal{K}_{z \bar z} (z, \bar z)$, and a choice of the potential $V_E(z, \bar z)$,   which define the bosonic part of the Einstein frame supergravity Lagrangian
\begin{eqnarray}
\mathcal{L}_{E}(z, \bar z)=\sqrt {-g_E}\left[  \frac{1}{2} {R}\left (g_E\right)
-\mathcal{K}_{z \bar z}  {\partial}_\mu z  {\partial}^\mu \bar z
-V_E (z, \bar z) \right]\, .
  \label{Turin-41E}
\end{eqnarray}
This Lagrangian is the Jordan-frame Lagrangian \rf{Turin-41} taken at 
\be
\Omega=1, \qquad g_{_J} \to g_{_E} \ .
\ee
But if in cosmological applications we are interested in models with a non-linear coupling of gravity to scalars, we need all 3 functions in eq. \rf{rule} and the action in eq. \rf{Turin-41}.
In cosmology, we often consider a theory with a single real scalar field, i.e., $z=\bar z$. We will therefore present the supergravity Lagrangians depending on a complex scalar at the one-dimensional slice $z=\bar z$, depending only on the real part of $z$.

\section{One-dimensional slice of the supergravity action}\label{Sec:1sc}

In cosmological applications, the relation between models in the Jordan frame with non-minimal coupling of multiple scalars to gravity and the Einstein frame was described in \cite {Salopek:1988qh}. More recently, in the context of $\xi$-attractors in \cite{Galante:2014ifa}, the Jordan frame action for one real scalar field was proposed in the form
\be\label{Jordan}
{\mathcal{L}_{\rm J}\over \sqrt{-g_J}} =  \frac12  \Omega( \phi) R  - \frac12 K_J(\phi ) (\partial \phi )^2 -  V_J(\phi ) \ .
\ee
and in the Einstein frame, the same theory was given as 
\be
 {\mathcal{L}_{\rm E}\over \sqrt{-g}} =  \frac12   R - \frac12 K_E(\phi ) (\partial \phi )^2    - V_E(\phi )    \ ,  
 \label{EinsteinG}
 \ee 
where
\be
K_E(\phi )={K_J(\phi ) \over \Omega(\phi )} + {3\over 2} {(\Omega')^2\over \Omega^{2}}, \quad 
   V_E(\phi ) =  \frac{V_J(\phi )}{\Omega^2(\phi )}  \ .
\label{KVE}
\ee
and 
\be
K_J= K_E\, \Omega -{3\over 2} {(\Omega')^2\over \Omega}\, , \quad 
 V_J=    V_E \, \Omega^2\  . 
\label{newKVJ} \ee 

We will now compare this construction with the supergravity actions in both frames, depending on the complex scalar $( \phi + i \chi)$ when we take it at the one-dimensional slice at $z=\phi $, $\chi=0$.  This will tell us how to embed the construction in \cite{Galante:2014ifa} in supergravity and what is the meaning of $K_E(\phi)$ and $K_J(\phi)$  as coming from $\mathcal{K}_{z \bar z} (z, \bar z)$ and from $(g_{_J})_{z \bar z} (z, \bar z)$, respectively.

We now consider the supergravity actions in the Einstein frame in Eq. \rf{Stanford} with one complex scalar 
\begin{eqnarray}
\mathcal{L}_{E} (\phi) = \sqrt {-g_E}\left[  \frac{1}{2} {R}\left (g_E\right)
-\mathcal{K}_{z \bar z} (z, \bar z)  {\partial}_\mu z  {\partial}^\mu \bar z
-V_E \right]\Big |_{z=\bar z=  \phi}\,.
  \label{Stanford1}
\end{eqnarray}
 We note that at $\chi=0 $ 
\be
\partial z \partial \bar z=  (\partial\phi)^2 \ .
\ee
We compare the slice of the supergravity action in eq. \rf{Stanford1} with the Einstein frame action for a scalar $\phi$ in \rf{EinsteinG}
 and find that these agree if 
 \be
{1\over 2} K_E(\phi ) = g_{z \bar z}|_{z=  \phi}= \mathcal{K}_{z \bar z}|_{z=\phi} =  {\frac{\partial^2  \mathcal{K}(z, \bar z)}{\partial z  \partial \bar z}}|_{z=  \phi} >0 \ . 
\label{KE} \ee
We consider the supergravity Jordan-frame action in Eq. \rf{Turin-41} and take its one-dimensional slice $z=  \phi$, $\chi=0$. With $\Omega(z, \bar z)= \Omega( \phi + i \chi, \phi - i \chi)$ we find that at $\chi=0$
\be
\Omega_z=\Omega_{\bar z} ={1\over 2} \Omega'= {1\over 2}{\partial \Omega (\phi)\over \partial \phi} \ .
\ee
Therefore
\be
(g_{_J})_{z \bar z} \partial z \partial \bar z = {1\over 2} \Big(\Omega \, K_E -{3\over 2}  {(\Omega')^2
\over \Omega }\Big)   (\partial\phi)^2 \ .
\label{gJ}\ee
One can also see from eq. \rf{A} that 
\be
{\cal A}^2_\mu(z, \bar z) |_{\chi=0}=0 \ .
\label{A11}\ee
Now compare the supergravity Jordan-frame action  \rf{Turin-4} at the one-dimensional slice using eqs. \rf{KE}, \rf{gJ} with  the Jordan frame action \rf{Jordan} used in \cite{Galante:2014ifa}: 
 These two actions coincide if
\be
\Omega(z, \bar z)|_{z=\bar z=  \phi}= \Omega(\phi)\, , \quad  g_{z \bar z}|_{z=\bar z=  \phi} ={1\over 2}  K_E(\phi)  \, , \quad 
 \Big ( g_{z\bar z} \Omega - 3\frac{\Omega
_z \Omega _{\bar z }}{\Omega }\Big )\Big|_{z=\bar z =\phi} ={1\over 2} K_J(\phi) \ .
\label{1d}\ee 
This means that the choice of the cosmological model with non-linear coupling of gravity to scalars begins with identifying the frame function and the \K potential in eq. \rf{two} and \K metric in eq.  \rf{Km} and their one-dimensional slice in \eqn{1d}.
Also the potential $V_E(z, \bar z)|_{z=\bar z= \phi}= V_E(\phi) $ has to be chosen. 

If the functions $\Omega(z, \bar z),   \mathcal{K}(z, \bar z), V_E (z, \bar z)$ are known, one can use the rules given above to get the Einstein and Jordan frame supergravities and their one-dimensional slices, which we need for cosmology.

The difference between the supergravity approach and the setup used in \cite{Kallosh:2013tua,Galante:2014ifa} is that in supergravity, we have to start with the fundamental function $\mathcal{K}(z, \bar z)$ which defines $K_E$, and proceed from there. This differs from the more traditional approach used in \cite{Kallosh:2013tua,Galante:2014ifa}, in which the models were based on a choice of $K_J$. 

The relation between $K_J$ and $K_E$ and $\Omega$, $\Omega'$ in eq. \rf{newKVJ} used in \cite{Galante:2014ifa} is compatible with supergravity if the conditions \rf{1d} are satisfied. So far, we have used the complex  fields in the form
$
z=  \phi + i \chi
$
to facilitate the relation between supergravity and standard Jordan-frame one-field models in eq. \rf{Jordan}.

\noindent We now  define new $\xi$-attractor action  as a Jordan-frame supergravity action
in the form given in Eq. \rf{Turin-41} where $z=T$ 
\begin{eqnarray} \boxed {
{\mathcal{L}_{J}\over \sqrt{-g}_J} (T, \overline{T})=\left[\Omega \left(  \frac{1}{2} {R}-3 {\cal A}^2_\mu(T, \overline{T}) \right )
-\left( \Omega g_{T\overline{T} }-3\frac{\Omega
_T \Omega _{\overline{T} }}{\Omega }\right)  {\partial}_\mu T  {\partial}^\mu \overline{T}
-\Omega^2 V_E \right] }
  \label{TurinJordan}
\end{eqnarray}
We use the $T$ variable here since in known attractor models in hyperbolic geometry, it is a half-plane variable with $T+\overline{T}>0$. Also, in polynomial attractor supergravity models in \cite{Kallosh:2025dac}, we have used the $(T, \bar {T})$ variables.
All models we will present below will be given in the form of \eqn{TurinJordan} either in the Jordan gauge\footnote{This choice of the frame function allows us to compare easily with the old $\xi$-attractors in\cite{Kallosh:2013tua,Galante:2014ifa} where for real scalars we had $\Omega= 1+\xi\phi^n$. Namely, we will find that keeping the same frame function but changing the choice of the Jordan kinetic term and Jordan potential, we can construct theories equivalent to earlier $\a$-attractors and general polynomial attractors in models with minimal coupling of scalars to gravity.}
 \be
\Omega(T, \overline{T})= 1+ \xi ( T \, \overline{T} )^{n\over 2} \ ,
\ee
or in the Einstein gauge
\be
\Omega(T, \overline{T})=1 \ .
\ee

\section{New  supergravity $\xi$-attractors} 
\subsection  {Hyperbolic $n=2$ geometry and $\xi={1\over 6\a}$}\label{Sec:new}

We start with reminding that there are different types of $\a$-attractors:  exponential E-models and  T-models  \cite{Kallosh:2013yoa},  polynomial $\a$ attractor models \cite{Kallosh:2022feu}, and singular  $\a$-attractors \cite{Kallosh:2025sji}. 

All  supergravity models for all $\alpha$-attractors   \cite{Kallosh:2013yoa,Kallosh:2022feu,Kallosh:2025sji} can be defined by the \K potential and \K metric 
\be
{\mathcal K}= -3\a \log (T+\overline{T}) \ , \qquad g_{T \overline{T}} = 3\a {1\over (T+\overline{T})^2} \ ,
\ee
where $T$ is a complex half-plane coordinate with ${\rm Re} \, T >0$.
The  action in the Einstein frame has the same hyperbolic geometry kinetic term for all these models, but the potentials are different  
\be
{ {\cal L}_E (T, \overline{T})\over \sqrt{-g}} =  {R\over 2} - {3\alpha} \, {\partial T \partial \overline{T}\over (T+\overline{T})^2}-  V_E(T, \overline{T}) \ .
\label{hyperE}\ee 
The kinetic term has an $SL(2, \mathbb{R})$ symmetry, but the potentials break this symmetry.
The potentials for exponential E-models and  T-models, and for polynomial $\a$-attractor models\footnote{We consider here for simplicity only the case of even $k$, for example, $k=2,4,6$. For more general $k$ we refer to  potentials in \cite{Kallosh:2022feu}.} are

\

\vskip - 1 cm
\be
V_E^{E\, model}(T, \overline{T}) =V_0  [ (1-T) (1-\overline{T})]^{m} \ ,
\label{VEE}\ee

\be
V_E^{T\, model} (T, \overline{T})= V_0 \left [ {1-T\over 1+T} \, { 1-\overline{T}\over 1+\overline{T}} \right ]^{m} \ ,
\label{VET}\ee

 \be
V^{polynomial}_E(T, \overline{T}) = V_0  { 1\over  1+ \Big (\ln^2 {T+\overline{T}\over 2}\Big )^{-m}}\, \label{Vpol}\ , 
\ee

\be
V^{sing}(T, \overline{T}) =V_E^{E ,T \, model}(T, \overline{T})  \Big (1+ \delta  \    \Big ({T+ \overline{T}\over 2}\Big)^{{-1}} \Big )  \, , \quad \delta \ll 1 \ .
\label{EsingT}\ee

We can switch from the geometric half-plane complex coordinate   $T$ to a dilaton-axion pair 
\be 
T=e^{-{\sqrt {2  \over 3 \alpha}} \varphi }+i\theta\, , \qquad     {\vp\over \sqrt 6\a}  ={1\over 2} \ln {T+\overline{T}\over 2} \ .
\ee

\noindent We now {\it  define new $\xi$-attractor action  as a Jordan-frame supergravity action}
in the form given in  \eqn{TurinJordan} where $T$ is a  complex half-plane coordinate with ${\rm Re} \, T >0$
with the choice 
\be
g_{T\overline{T} }=  {1\over 2\xi} {1\over (T+\overline{T})^2}  \ .
\label{metric}\ee
Using the superconformal symmetry in eq. \rf{TurinJordan} we can choose  the frame function depending on $(T, \overline{T})$ 
 \be
{\rm Jordan \, \, gauge} \, \,  :  \qquad \Omega(T, \overline{T})= 1+ \xi \, T \, \overline{T}  \ ,
\label{OT1}\ee
and find other ingredients of the Jordan frame action \be
 \qquad  \Omega_T = \xi \overline{T} ,  \quad  \Omega_{\overline{T}} = \xi  T \, , \quad {\cal A}_\mu(T,{\overline{T}}) \equiv
-\frac{\rmi}{2\Omega }\,
  \left(  {\partial}_\mu T\, \Omega_T
  -  {\partial} _\mu \overline{T}\, \Omega_{\overline{T}} \right) \ .
\label{OT2}\ee
With this choice, using the action in \rf{TurinJordan}, we have a Jordan frame action with field variables $T, \overline{T}$. 
 We will present  here the Jordan frame action of new $\xi$-attractors in Jordan gauge  at $T=\overline{T}$, where 
\be
\Omega(T, \overline{T})\to 1+ \xi \, T ^2\, \, , \quad \Omega_T \to \xi T ,  \quad  \Omega_{\overline{T}} \to \xi T\, ,\quad A_\mu\to0\, , \quad g_{T\overline{T}} \to {1\over 8\, \xi \, T^2} \ ,
\ee
\begin{eqnarray}
{\mathcal{L}_{J}\over \sqrt{-g}_J}|_{T=\overline{T}} = (1+ \xi \, T ^2)   \frac{1}{2} {R} 
-\left( {(1+ \xi \, T ^2)
\over 8\, \xi \, T^2} -3 \frac{ \xi^2  T^2}{(1+ \xi \, T^2 )}\right) ( {\partial} T)^2   
- V_J ( T) \ .
  \label{TurinE1}
\end{eqnarray}
where $
V_J (T)\to (1+ \xi \, T ^2)^2 V_E (T)$
 and 
\be
V_E^{E\, model}(T) =V_0 [ (1-T) ]^{2m}, \quad V_E^{T\, model} (\vp)= V_0 \left [ {1-T\over 1+T} \right ]^{2m}, \quad V^{polynomial}_E(\vp) ={ V_0 \over  1+ \left(\log T \right)^{-2m}} 
\label{VTr}\ee
\noindent  Using the superconformal symmetry in eq. \rf{TurinJordan} we can choose  the frame function independent  on $(T, \bar T)$ 
\be
{\rm Einstein \, \, gauge} \, \,  :  \qquad \Omega^E(T, \bar T)= 1
\label{OT3}\ee
and find other ingredients of the Einstein frame action
$
 \Omega^E_T = 0$ ,  $ \Omega^E_{\overline{T}} =0$ , $ {\cal A}^E_\mu(T, {\overline{T}}) =0$.
We get  an Einstein frame action in the form
\be
{ {\cal L}_E (T, \overline{T})\over \sqrt{-g}} =  {R\over 2} - {1\over 2 \,\xi} \, {\partial T \partial \overline{T}\over (T+\overline{T})^2}-  V_E(T, \overline{T}) \ .
\label{hyperEJ}\ee  
At the one-dimensional slice $T=\overline{T} = e^{- 2 {\sqrt \xi}  \varphi }$ we find the potentials $V(\vp)$ in \rf{VTr},  respectively to be
\be
V_0 \left ( 1-e^{- 2 {\sqrt \xi}  \varphi } \right )^{2m}, \qquad  V_0 \left ( \tanh{{\sqrt \xi} \varphi } \right )^{2m}, \qquad V_0 \Bigg({ 1\over  1+ \left  ( {2 \sqrt \xi \, \vp} \right)^{-2m}}\Bigg ) \ .
\label{canV}\ee
Comparing with the relevant potentials in $\a$-attractors
\be
V_0 \left ( 1-e^{-  {\sqrt {2\over 3 \a} }  \varphi } \right )^{2m}, \qquad V_0 \left ( \tanh{1\over {\sqrt 6\a} \varphi } \right )^{2m} , \qquad V_0 \Bigg({ 1\over  1+ \left  (  {\sqrt {2\over 3 \a} \varphi} \right)^{-2m}}\Bigg )\ .
\ee
It is easy to see that the theory of new $\xi$-attractors in the Einstein gauge in the form \rf{hyperEJ} coincides with the $\a$-attractors  in eq. \rf{hyperE} under condition that
\be
\xi ={1\over 6\a}  \ .
\ee
In the polynomial case, they are equivalent to KKLTI models with $k=2m$.

\subsection{General polynomial $\xi, n\neq 2$ attractors }\label{Sec:pol}
We start with a supergravity version of the KKLTI model \cite{Martin:2013tda,Kallosh:2018zsi,Kallosh:2019hzo}  depending on $(\mu, k)$.
We choose the \K potential closely related to the one in  \cite{Kallosh:2025dac} 
\be
K(T, \overline{T}) =   {\mu^2  \over  2 ( {T} \overline{T} )^{1\over k}} \ .
\label{Kkk} \ee
 The \K metric follows
\be
g_{T \overline{T}} = {  \mu^2\over  2 k^2}   {1\over ( {T} \overline{T} )^{k+1\over k}} \ ,
\label{kkm}\ee
and the Einstein frame Lagrangian is
\be
{ {\cal L} (T, \overline{T})\over \sqrt{-g}} =  {R\over 2} - {  \mu^2\over 2 k^2}   {\partial T \partial \overline{T}\over ( {T} \overline{T} )^{k+1\over k}}
-  V_E(T, \overline{T})    \ .
\label{hyperEP}\ee 
Here the $SL(2, \mathbb{R})$ symmetry of the kinetic term is broken.
We can relate the field $T$ to $\vp$ and $\theta$ as follows
\be
T= \left ({ \mu^2\over \vp^2 }\right)^{k\over 2}e^{i\theta} \, ,  \qquad {T \overline{T}}  = \Big({ \mu^2\over \vp^2 }\Big )^k \ ,
\ee
 where $g_{\vp\vp} =1/2$, $\vp$  is a canonical field. The KKLTI potential at even $k$, for example, $k=2,4,6$, is defined as follows
\be
V_{KKLTI}(\vp)= { 1\over 1+\Big({ \mu\over \vp }\Big )^{k} } \ .
\ee
We may relax the restriction to  even $k$ by taking 
\be
V_E(T, \overline{T}) = { 1\over  1+(T\overline{T})^{1\over 2} }\,  \qquad\rightarrow \qquad V_E(\vp)= { 1\over 1+\Big({ \mu^2\over \vp^2 }\Big )^{k\over 2} } \ .
\label{Vn}\ee
\noindent We now {\it  define new $\xi$-attractor models depending on $(\xi, n)$   as a Jordan-frame supergravity action}
in the form given in  \eqn{TurinJordan}. The \K potential ${  \mu^2\over 2 \, ( {T} \overline{T} )^{1\over k}}$ will be given in a form where $k$ is a function of $n$  where we consider  two cases, $n>2$ and $n<2$.
\be
k={2\over n-2}\, , \quad n>2 \ ,
\ee
or 
\be
k={2n\over 2-n}\, , \quad n<2 \ .
\ee
In both cases, we will  define the Jordan frame theory using
\be
 \Omega= 1+\xi (T\overline{T})^{n\over 2} \ ,   \qquad \xi= {k^2\over \mu^2} \ .
\ee

\noindent {\bf New $\xi$-attractors with $n>2$ }

\noindent Here we replace ${k+1\over k}$ in eq.  \rf{kkm} by ${n\over 2}$ and $\mu^2/k^2$ by $ {1\over \xi}$ and 
the \K metric follows
\be
g_{T \overline{T}} = {  1\over 2 \xi}   {1\over ( {T} \overline{T} )^{n\over 2}} \ .
\ee
We may choose the frame function $\Omega (T, \overline{T}) $  as 
\be
 {\rm Jordan \, \, gauge} \, \,  :  \qquad \Omega= 1+\xi (T\overline{T})^{n\over 2}  \ ,
\ee
where 
  \be\label{xikmuw}
 k={2\over n-2}  \to  \quad n= {2(k+1)\over k} \ .
 \ee
The Jordan frame supergravity action takes the form given in eq. \rf{TurinJordan} with   $\Omega= 1+\xi (T\overline{T})^{n\over 2}$,  $g_{T \overline{T}} = {  1\over 2 \xi}   {1\over ( {T} \overline{T} )^{n\over 2}}$ and the potential $V_0{\Omega^2\over   1+(T\overline{T})^{1\over 2} }$.
At the one-dimensional slice $T=\overline{T}$ we get 
 \be\label{JordanPolcan1}
{\mathcal{L}_{\rm J}\over \sqrt{-g_J}} =  \frac12  (1+\xi T^n) R  -  \Big (   {1+\xi T^n\over 2 \xi ( {T} )^{n}}
 -{3\over 4} {n^2 \xi^2 T^{(n-2)}\over 1+\xi T^n}\Big) (\partial T )^2-  V_0{ (1+\xi T^n) ^2\over 1+T  } \,  \ .
\ee
The Einstein frame Lagrangian is given by the general expression  in \rf{TurinJordan} with the choice 
\be
{\rm Einstein \, \, gauge} :  \qquad \Omega^E(T, \bar T)= 1
\label{OT4}\ee
\be
{\mathcal{L}_{\rm E}\over \sqrt{-g}} =  \frac12   R  - {  1\over 2 \xi}   {\partial T \partial \overline{T}\over ( {T} \overline{T} )^{n\over 2}}  -   { V_0\over 1+(T\overline{T})^{1\over 2} }\ .
\label{canG}\ee 
where we can use 
\be
T= \left ({ \mu^2\over \vp^2 }\right)^{1\over n-2}e^{i\theta} \, ,  \qquad {T \overline{T}}  = \Big({ \mu^2\over \vp^2 }\Big )^{2\over n-2} \ .
\ee
At the one-dimensional slice $T=\overline{T} =  \left ({\mu^2 \over \vp^2}\right)^{1\over n-2}$ we find the potentials $V(\vp)$ in \rf{Vn},   to be
\be
{ V_0\over 1+(T\overline{T})^{1\over 2} } = { V_0\over 1+ \left ({ \mu^2\over \vp^2 }\right)^{1\over n-2} } \ .
\label{V1}\ee
Thus, new polynomial  $\xi$-attractors with non-minimal coupling of a scalar  to gravity $\xi T^n$ and $n>2$ are equivalent to a general KKLTI-type attractor \cite{Martin:2013tda,Kallosh:2018zsi,Kallosh:2019hzo} 
 defined by $(\mu, k)$ under condition that
 \be
k={2\over n-2} \ , \quad \mu^2 =  {k^2 \over \xi} \, , \quad n= {2(k+1)\over k} >2 \ .
 \ee

 \noindent {\bf New $\xi$-attractors with $n<2$ }
 
  \noindent Here we replace ${k+1\over k}$ in eq.  \rf{kkm} by ${n+2\over 2n}$ and $\mu^2/k^2$ by $ {1\over \xi}$, and the \K metric follows
\be
g_{T\overline{T} }= {  1\over 2 \xi}   {1\over ( {T} \overline{T} )^{n+2\over 2n}} \ .
\ee
We may choose the frame function $\Omega (T, \overline{T}) $  as 
\be
 {\rm Jordan \, \, gauge} \, \,  :  \qquad \Omega= 1+\xi (T\overline{T})^{n\over 2} \ ,
\ee
where 
  \be\label{xikmuw1}
k={2n\over 2-n}   \to  \quad n={2k\over 2+k} \ .
 \ee
 The Jordan frame supergravity action takes the form given in eq. \rf{TurinJordan} with   $\Omega= 1+\xi (T\overline{T})^{n\over 2}$,  $g_{T \overline{T}} = {  1\over 2 \xi}   {1\over ( {T} \overline{T} )^{n+2\over 2n}}
$ and the potential $V_0{\Omega^2\over   1+(T\overline{T})^{1\over 2} }$.
At the one-dimensional slice $T=\overline{T}$ we get

 \be\label{JordanPolcan}
{\mathcal{L}_{\rm J}\over \sqrt{-g_J}} =  \frac12  (1+\xi T^n) R  - \frac12 \Big (   {1+\xi T^n\over 2 \xi ( {T} )^{n+2\over n}}
 -{3\over 4} {n^2 \xi^2 T^{(n-2)}\over 1+\xi T^n}\Big) (\partial T )^2-  V_0{ (1+\xi T^n) ^2\over 1+T  } \,  \ .
\ee
The Einstein frame Lagrangian is given by the general expression  in \rf{TurinJordan} with the choice 
\be
{\rm Einstein \, \, gauge} :  \qquad \Omega^E(T, \bar T)= 1
\label{OT5}\ee
\be
{\mathcal{L}_{\rm E}\over \sqrt{-g}} =  \frac12   R  - {  1\over 2 \xi}   {\partial T \partial \overline{T}\over ( {T} \overline{T} )^{n+2\over n} } -   { V_0\over 1+(T\overline{T})^{1\over 2} }\ .
\label{canG1}\ee 
where we can use 
\be
T= \left ({ \mu^2\over \vp^2 }\right)^{n\over 2-n}e^{i\theta} \, ,  \qquad {T \overline{T}}  = \Big({ \mu^2\over \vp^2 }\Big )^{2n\over n-2} \ .
\ee
At the one-dimensional slice $T=\overline{T} =  \left ({\mu^2 \over \vp^2}\right)^{n\over n-2}$ we find the potentials $V(\vp)$ in \rf{Vn},   to be
\be
{ V_0\over 1+(T\overline{T})^{1\over 2} } = { V_0\over 1+ \left ({ \mu^2\over \vp^2 }\right)^{n\over 2-n} } \ .
\label{V2}\ee
Thus, new polynomial  $\xi$-attractors with non-minimal coupling of a scalar  to gravity $\xi T^n$ and $n<2$ are equivalent to a general KKLTI-type attractors \cite{Martin:2013tda,Kallosh:2018zsi,Kallosh:2019hzo} 
 defined by $(\mu, k)$ under condition that
 \be
k={2n\over 2-n}  \ , \quad \mu^2 =  {k^2 \over \xi} \, , \quad n={2k\over 2+k} < 2 \ .
 \ee 
\subsection{Universal form of new supergravity $\xi$-attractors}

Superconformal theory upon gauge-fixing local Weyl symmetry, special conformal symmetry, and special supersymmetry defines  Poincar\'e supergravity \cite{Kallosh:2000ve,Ferrara:2010yw,Ferrara:2010in,Freedman:2012zz}. When the nilpotent supermultiplet is present, these theories are defined by three independent functions shown in Eq. \rf{rule}, the frame function $\Omega(z, \bar z)$, the \K potential $\mathcal{K} (z, \bar z)$ and the potential $V_J(z, \bar z)= \Omega^2 V_E (z, \bar z) $.

We have presented here the bosonic part of the supergravity action in a Jordan frame with non-minimal coupling of scalars to gravity 
\be
\Omega=1+\xi (T\overline{T})^{n/2} \ .
\ee
 The supergravity action for all exponential and polynomial $\xi$-attractors has a universal form. We present here its scalar-gravity part
\begin{eqnarray}
{\mathcal{L}_{J}\over \sqrt{-g}_J} =\left[\Omega \left(  \frac{1}{2} {R}-3 {\cal A}^2_\mu(T, \overline{T}) \right )
-(g_{_J})_{T \overline{T}}  {\partial}_\mu T  {\partial}^\mu \overline{T}
-\Omega^2 V_E \right]\ ,
  \label{TurinEpol}
\end{eqnarray}
where 
\be
(g_{_J})_{T \overline{T}} = \Omega \, \mathcal{K}_{T \overline{T}} -3 \frac{\Omega
_T \Omega _{\overline{T} }}{\Omega }\, ,\qquad {\cal A}_\mu(T,{\overline{T}}) \equiv
-\frac{\rmi}{2\Omega }\,
  \left(  {\partial}_\mu T \, \Omega_T
  -  {\partial} _\mu \overline{T}\, \Omega_{\overline{T}} \right) .
\ee
The three $n=2$ models associated with hyperbolic geometry, exponential T and E \, $\xi$-attractors, and polynomial  $\xi$-attractors  have the following \K potential and frame function
 \be
{\mathcal K}= -{1\over 2 \, \xi}  \log (T+\overline{T})\, , \qquad 
\Omega(T, \overline{T})= 1+ \xi \, T \, \overline{T} \ ,
\ee
which leads to a non-minimal coupling to gravity $\xi T^2 R$.
The potentials $V_E$  are different for these three models, T, E, and polynomial, respectively
\be
[ (1-T) (1-\overline{T})]^{m}\, , \quad 
\left [ {1-T\over 1+T} \ { 1-\overline{T}\over 1+\overline{T}} \right ]^{m}\, , \quad 
{ 1\over  1+ \Big (\ln^2 {T+\overline{T}\over 2}\Big )^{-m}}  \ .  \label{Vpol1}\ee
The general polynomial $\xi$-attractors with a  non-minimal coupling to gravity  have 
\be
 \Omega=  1+\xi (T\overline{T})^{n\over 2} \, , \quad V_E(T, \overline{T}) = { 1\over  1+({T\overline{T}})^{1/2} } \ .
\ee
In case $n>2$ 
\be
K(T, \overline{T}) =  {  1\over 2 \xi \, ( {T} \overline{T} )^{n-2\over 2}} \left({2\over 2-n}\right)^2\ ,  
\ee
In case $n<2$ 
\be
K(T, \overline{T}) =  {  1\over 2 \xi \, ( {T} \overline{T} )^{2-n\over 2n}} \left({2n\over 2-n}\right)^2\, .  
\ee
In the gauge $\Omega^E=1$ and at the slice $T =\overline{T}$, all these models can be shown to reduce to a Lagrangian with a canonical field $\vp$ 
\be\label {Ein4}
{\mathcal{L}_{\rm E}\over \sqrt{-g}} =  \frac12   R  - \frac12  (\partial \vp )^2 -  V_E(\vp) \ .
\ee 
where the corresponding three potentials in hyperbolic models are given in eq. \rf{canV}
and the potential in general polynomial $\xi$-attractors  is given in eq. \rf{V1} for $n>2$ and in eq. \rf{V2} for $n<2$.

A more detailed discussion of the cosmological consequences of the new $\xi$ attractors will appear in \cite{KallLin:2026nn}.

\section{ Discussion}
Jordan and Einstein frames in supergravity interacting with chiral multiplets are given by two different gauges of the  $SU(2,2|1)$ superconformal theory, where the local Weyl symmetry, special conformal symmetry, and special supersymmetry are gauge-fixed \cite{Kallosh:2000ve,Ferrara:2010yw,Ferrara:2010in,Freedman:2012zz}.

In this paper, we have constructed a supergravity version of the new cosmological $\xi$-attractors depending on the choice of the \K potential 
 $ \mathcal{K}$ and the frame function $\Omega$ 
 \be
 \mathcal{K}(z, \bar z)  \, , \qquad \Omega(z, \bar z)  \ .
\label{two1}\ee 
Einstein frame supergravity kinetic term of scalars is defined by $ \mathcal{K}(z, \bar z)$ and there is no scalar coupling to gravity, $ \Omega(z, \bar z)=1$. In Jordan frame supergravity, $\Omega(z, \bar z)$ defines the coupling of scalars to gravity, and both $ \mathcal{K}(z, \bar z)$ and $\Omega$ define the kinetic term of scalars. These two frames in supergravity originate from gauge-fixing superconformal theory in one or another gauge, a scalar independent $\Omega(z, \bar z)=1$ or a specific choice of a function of scalars, $\Omega(z, \bar z)$. Supergravities in these two gauges are related by a specific local superconformal transformation that includes a local Weyl transformation, which is used in a theory of a single scalar field with minimal or non-minimal coupling to gravity.

We presented supergravity versions in both the Jordan and Einstein frames. 
We gave examples of exponential and polynomial new $\xi$-attractors, both in the Jordan frame with non-minimal coupling of scalars to gravity and in the Einstein frame with minimal coupling of scalars to gravity. 

The supergravity kinetic term for the complex scalars in the Einstein  frame is
\be
 g_{z \bar z}\partial z \partial \bar z\, , \qquad {\rm where} \quad  g_{z \bar z}= {\frac{\partial^2  \mathcal{K}(z, \bar z)}{\partial z  \partial \bar z}}\equiv \mathcal{K}_{z \bar z} \ .
\ee
The supergravity kinetic term for the complex scalars in the Jordan frame with the frame function  $\Omega(z,\bar z)$  is
\be 
(g_{_J})_{z \bar z}\partial z \partial \bar z\, , \qquad {\rm where} \quad (g_{_J})_{z \bar z} = \Omega \, \mathcal{K}_{z \bar z} -3 \frac{\Omega_z \Omega _{\bar z }}{\Omega } \ .
\ee
There are also additional kinetic terms in the Jordan frame that depend on the square of the auxiliary vector $\mathcal{A}_\mu$; see eqs. \rf{Turin-4}, \rf{A}. These terms vanish at $z=\bar z$.

We have shown that the supergravity relation between the moduli space metrics $(g_{_J})_{z \bar z}$ and $g_{z \bar z}$ taken at $z=\bar z=\phi$ is precisely the relation between $K_J$ and $K_E$ which was suggested and used in \cite{Galante:2014ifa}
\be
[(g_{_J})_{z \bar z} = \Omega \, \mathcal{K}_{z \bar z} -3 \frac{\Omega_z \Omega _{\bar z }}{\Omega }]\big | _{z=\phi } \quad \to 
 \qquad K_J(\phi)= K_E\, \Omega -{3\over 2} {(\Omega')^2\over \Omega}\, .
\ee
Supergravity  choice of cosmological models starts with the choice of the  \K potential and the relevant $K_E(\phi)$, and subsequent $K_J(\phi)$
\be
{\mathcal K}(z,\bar z)\,  \quad \to \quad g_{z \bar z} \,  \qquad \to \quad K_E(\phi) \,  \qquad \to \quad K_J(\phi) \ .
\label{1}\ee
This is an order opposite to the one in  \cite{Kallosh:2013tua,Galante:2014ifa}, which was 
\be
K_J \,  \quad \to \quad K_E \ .
\label{2}\ee
 In  \cite{Galante:2014ifa}  a specific  choice $K_J  \to  K_E$  led to a choice of $\xi$-attractors relation to $\a$-attractors under condition that $\a=1+{1\over 6\xi}$. Instead, using the process in eq. \rf{1} we came up, naturally, to a choice of $\xi$-attractors relation to exponential and polynomial $\a$-attractors under condition that 
\be
\a={1\over 6\xi} \ .
\ee
In this case, the limit of strong coupling $\xi\to \infty$ is a limit to small $\a\to 0$ and infinite \K curvature of the hyperbolic geometry. This regime  is particularly important with regard to a  future detection of primordial gravitational waves, where the attractor properties of the cosmological models predict the level of gravity waves as $r\approx {12 \a\over N^2}$, and as we have learned here
\be
r\approx {12 \a\over N^2}= {2\over \xi \, N^2} \ .
\ee
For $\a$-attractors, the detection of gravity waves means the measurement of the \K curvature of the hyperbolic geometry ${\cal R}_{\cal K}=-{2\over 3\a}$. Now we have learned that the detection of gravity waves means the measurement of $\xi$, the non-minimal coupling to gravity in new $\xi$-attractors defined here.

Finally, we can formulate the main property of the new $\xi$-attractors versus the old ones in \cite{Kallosh:2013tua,Galante:2014ifa}. For example, in \cite{Kallosh:2013tua} we have   the action
\be
{{\cal L}^J\over \sqrt{-g^J}}=  (1+\xi \phi^n) R - {1\over 2}  (\partial \phi)^2 -  V_0 \phi^{2n} \ .
\ee
\begin{figure}[H]
\vspace{-1mm}
\hspace{-3mm}
\begin{center}
\includegraphics[scale=0.4]{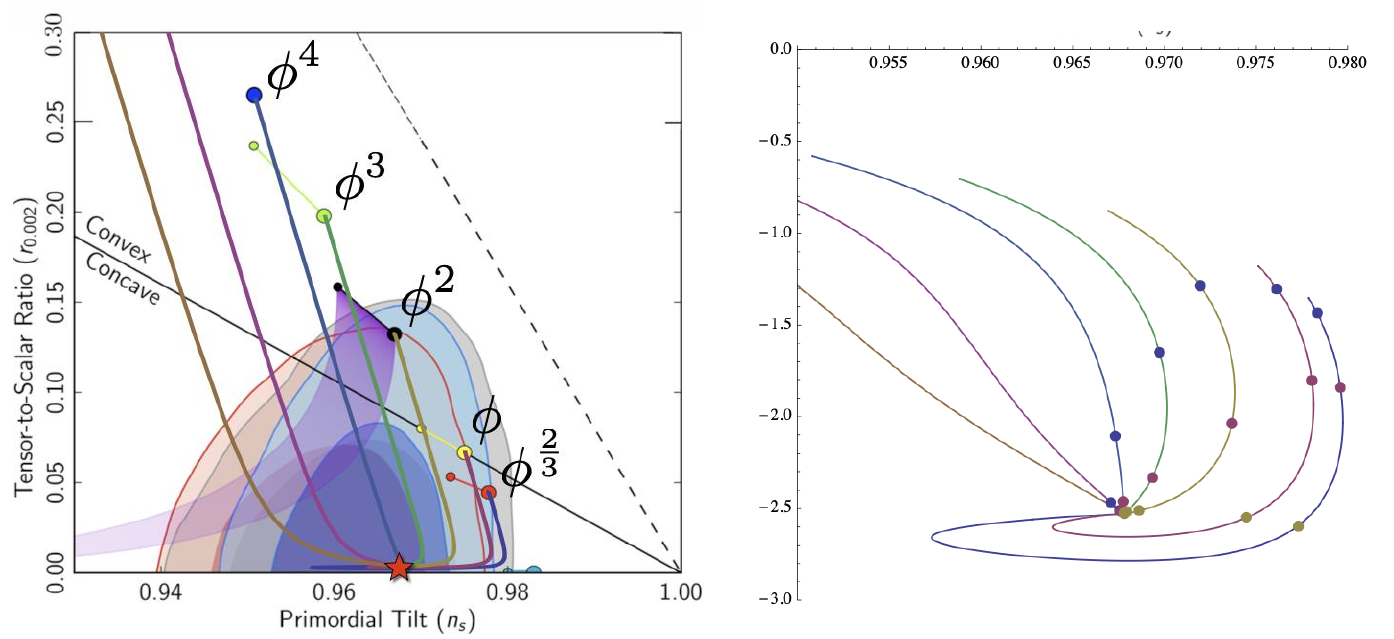} 
\end{center}
\vspace{-.12cm}
\caption{\footnotesize  Here we plot the values of $n_s, r$ for old $\xi$-attractors in   \cite{Kallosh:2013tua} on a linear and a
logarithmic scale
for  $n=2/3, 1,2,3,4,6,8$.  The values of $\xi$ go down from 0 to infinity along the trajectories with $r\geq {12\over N^2}$.}
\label{Old}
\end{figure}
For all $n$, the theory at strong coupling $\xi\to \infty$ with $\xi>0$ in both models developed in \cite{Kallosh:2013tua,Galante:2014ifa} 
 has the following prediction for $r$ illustrated by in Fig. \ref{Old}
\be
\xi\to \infty  :   \, \qquad  n_{s} = 1-{2\over N} \ , \qquad  r = {12 \over  N^{2}} \ .
\label{aa}\ee
\begin{figure}[H]
\vspace{-1mm}
\hspace{-3mm}
\begin{center}
\includegraphics[scale=0.22]{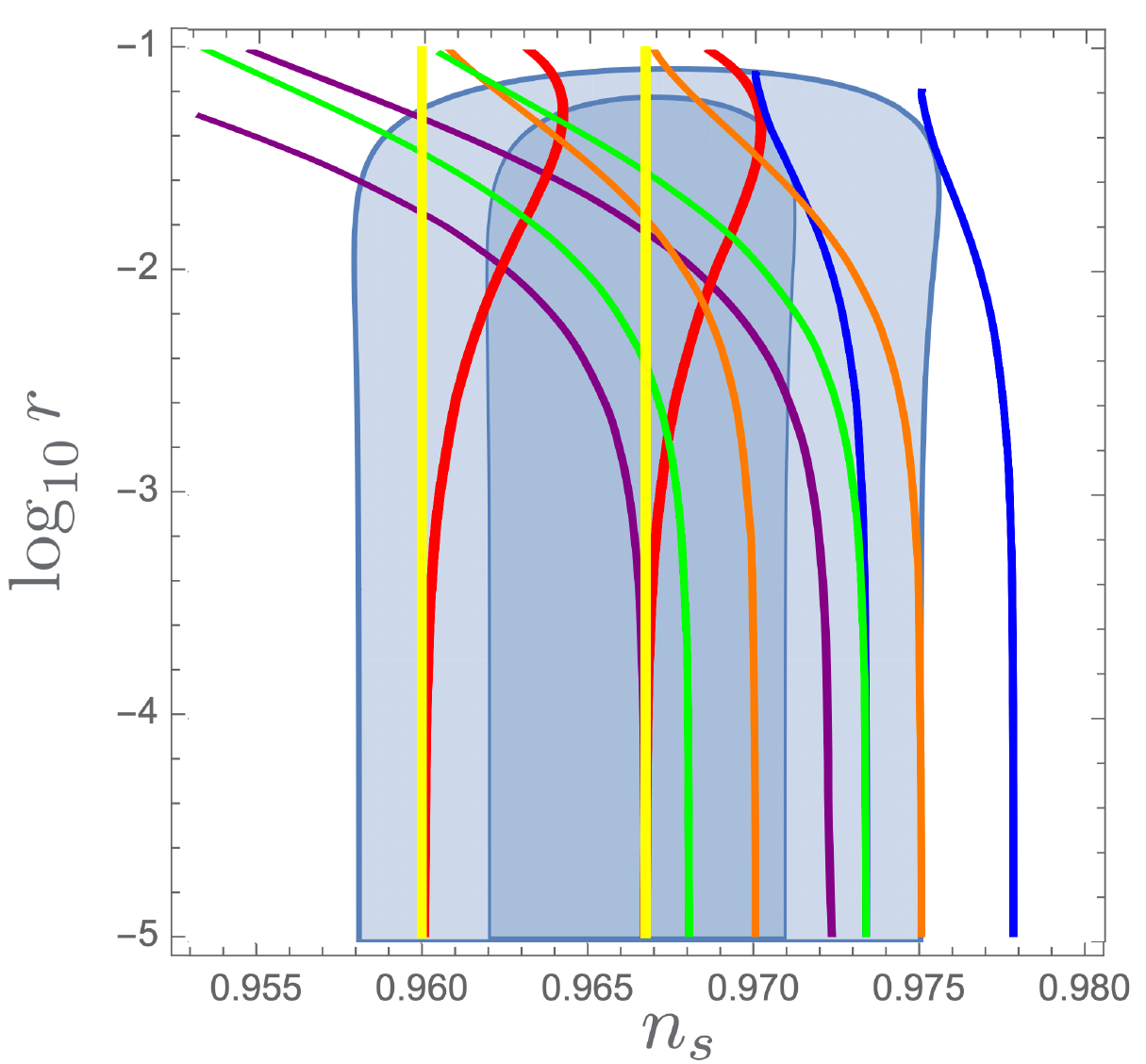} 
\end{center}
\vspace{-.12cm}
\caption{\footnotesize  New $\xi$-attractors T and E models, yellow and red lines,  with $\xi={1\over 6\a}$, 
and $k=4,3,2,1$ polynomial models, purple, green, orange, and blue lines, respectively. These have $\xi={1\over 6\xi}$, $n=2$ and $\mu^2={3\a\over 2}$ or $n\neq 2$ and $\xi={4\over (n-2)^2 \mu^2}$. The values of $\xi$ go down from 0 to infinity along the trajectories with $r\geq 0$.
}
\label{Blue}
\end{figure}
New $\xi$-attractors in this paper have the same first term, $(1+\xi \phi^n) R$, but different choice of $K_J,  V_J$ and have the property that 
\be
\xi\to \infty  :  \, \qquad  n_{s} \geq 1-{2\over N} \ , \qquad   r  \to 0 \ .
\label{aaa}\ee

This is the same behavior as in all cosmological attractors in minimal-coupling-to-gravity models \cite{Kallosh:2019hzo}, with the choice of such $K_J,  V_J$ supplied by the supergravity models based on superconformal theory. The reason is that cosmological attractors in the Einstein frame and in the Jordan frame are two different, classically equivalent, gauges of the superconformal theory.

 \section*{Acknowledgement}
We are grateful to  S. Ferrara, L. Kofman,  A. Linde, A. Marrani,  and A. Van Proeyen for a collaboration on superconformal theory, supergravity, and cosmology in the past, and to   A. Linde, D. Roest,  T. Terada,  and Y. Yamada for recent stimulating discussions.  This work is supported by the Leinweber Institute for Theoretical Physics at Stanford and by NSF Grant PHY-2310429

\appendix 
\section{Early $\xi$-attractors in supergravity}\label{A:Hxi} 
\noindent In \cite{Kallosh:2013tua} the supergravity version of $\xi$-attractors was defined by the following \K potential and the frame function (at $S=0$, where $S$ is a heavy stabilizer or a  nilpotent superfield) 
\be \Omega (\Phi, \bar \Phi) = 1+ {\xi\over 2} \left (f(\sqrt {2}\, \Phi)+ \bar f (\sqrt {2} \, \bar \Phi)\right )+ {1\over 6} (\Phi-\bar \Phi)^2, \quad 
{\mathcal K} (\Phi, \bar \Phi) = -3 \log \left (\Omega(\Phi, \bar \Phi)  \right) \ , \ee
\be
V_E= {f(\sqrt {2}\, \Phi) \bar f (\sqrt {2} \, \bar \Phi)\over \Omega^2(\Phi, \bar \Phi) } \ .
\ee
With this choice, the Einstein frame has a rather complicated kinetic term with $g_{\Phi \bar \Phi} = {\frac{\partial^2  \mathcal{K}(\Phi, \bar\Phi)}{\partial \Phi  \partial \bar \Phi}}$. The Jordan frame metric 
\be 
(g_{_J})_{\Phi \bar \Phi} = \Omega \, g_{\Phi \bar \Phi} -3 \frac{\Omega
_\Phi \Omega _{\bar \Phi }}{\Omega }\,   \label{gJA} \ee
is also a complicated function for the complex $\Phi$, however, at the one-dimensional slice $ (\sqrt {2}\, \Phi)= \phi$ the Jordan frame kinetic term  collapses to 
$
{1\over 2} (\partial \phi)^2 
$
in agreement with the original choice $K_J=1$ for this model. The  one-dimensional slice $ (\sqrt {2}\, \Phi)= \phi$ of the Einstein  frame kinetic term is still complicated, $K_E=  \Omega^{-1} +{3\over 2} {(\Omega')^2\over \Omega^2} $. The cosmological predictions of this model, including $n_s, r$, were obtained numerically and plotted in Fig. 1 in \cite{Kallosh:2013tua}. But it was also noticed there that at the strong coupling limit  $\xi\to \infty$ one finds that 
$K_E= \Omega^{-1} +  {3\over 2} {(\Omega')^2\over \Omega^2} \to {3\over 2} {(\Omega')^2\over \Omega^2}$ the action can be reduced to a simple one, for example, the one in eq. (15) in \cite{Kallosh:2013tua}
\be
{{\mathcal L}_E\over \sqrt{-g}}|_{\xi \to \infty} \to  {1\over 2}R -{1\over 2} (\partial \vp)^2- V_0 \left (1-\exp^{-\sqrt {2\over 3}\, \vp} \right)^2 \ ,
\ee
which one can qualify as an $\a$ attractor E model at $\a=1$. This explains why the strong coupling limit  $\xi\to \infty$  stops at $\a=1$. 
An analogous feature one finds in the $\xi$-attractors in \cite{Galante:2014ifa}, where \footnote{It is not clear how to promote the one  real scalar model of special attractors in \cite{Galante:2014ifa} to supergravity level}
\be
\alpha= 1 +{1\over 6\xi}\, , \qquad \a |_{\xi \to \infty} \to 1 \ .
\ee

 \section{Palatini formalism in supergravity and cosmology}\label{A:Pal}

 There was recently some interest in a first-order \'a la Cartan conformal gravity and supergravity, see for example
\cite{Francois:2024rfh,Cattaneo:2026onm,Andrianopoli:2026qsx}.  First, we explain why the constructions in \cite{Francois:2024rfh,Cattaneo:2026onm,Andrianopoli:2026qsx} are not relevant to Jordan frame supergravity.

The papers \cite{Francois:2024rfh,Cattaneo:2026onm,Andrianopoli:2026qsx} and references therein describe  Palatini-Cartan formalism in pure (no matter) $\cN = 1$, $D = 4$  supergravity. In particular, in  \cite{Francois:2024rfh} we find that Palatini-Cartan supergeometry constructions are closely related to MacDowell-Mansouri formulation of pure supergravity \cite{MacDowell:1977jt}. In the absence of matter, it is known that in \cite{MacDowell:1977jt} the first-order formalism, where spin connection $\omega_\mu^{ab}$ is an independent field,  leads to an equation of motion where the constraint making spin connection a function of the vierbein 
is recovered as an equation of motion. Therefore, the second order formalism has $\omega_\mu^{ab}$ as a function of the vierbein and a gravitino, as it is also well known in pure supergravity  \cite{Deser:1976eh}.

Here, we are interested in the properties of the Jordan frame supergravity where cosmological Palatini attractors are defined in models with a non-minimal coupling of gravity to scalars \cite{Tenkanen:2020dge}, starting with an arbitrary frame function $\Omega$, arbitrary $K_J$ and $V_J$. See also the most recent article on cosmology in Palatini gravity \cite{Bostan:2026ltp}  with  $K_J=1$. 

Pure supergravity Palatini-Cartan models in \cite{Francois:2024rfh,Cattaneo:2026onm,Andrianopoli:2026qsx}  have no physical scalar multiplets and therefore only one frame, the Einstein frame with 
\be
\Omega=1
\ee  
Thus, supergravity constructions \cite{Francois:2024rfh,Cattaneo:2026onm,Andrianopoli:2026qsx}  based on Palatini-Cartan supergeometry do not have a Jordan frame different from the Einstein frame and have no scalar fields to drive inflation.

Is it possible to have a supergravity theory of the kind used in Palatini formalism application to cosmology? A brief statement of the difference between the standard and Palatini formalisms is as follows, according to \cite{Tenkanen:2020dge}. The metric $g_{\mu\nu}$ and Christoffel connections $\Gamma^\mu_{\nu\lambda}$ are independent in Palatini case, i. e. Christoffel connections are not metric-compatible. In standard general relativity, metric-compatible Christoffel connections depend on the metric as follows
\be
D_\rho g_{\mu\nu}=0 \qquad \rightarrow \qquad \Gamma^\sigma_{\mu \nu }(g)={1\over 2}g^{\sigma \rho}( \partial_\mu g_{\nu\rho}+\partial_\nu g_{\mu\rho}-\partial_\rho g_{\mu\nu}) \ .
\label{comp}\ee
In both cases, standard and Palatini gravity,  the Ricci tensor depends on the Christoffel connection $\Gamma^\mu_{\nu\lambda}$, namely
\be
R_{\sigma\nu} = R^\lambda{}_{\sigma\lambda \nu} \ ,
\label{RP}\ee
where the Riemann-Christoffel 4-tensor is
\be
R^\rho{}_{\sigma \mu \nu}=\partial_\mu \Gamma^\rho_{\nu\sigma}- \partial_\nu \Gamma^\rho_{\mu\sigma}+\Gamma^\rho_{\mu\lambda}\Gamma^\lambda_{\nu\sigma}-\Gamma^\rho_{\nu\lambda}\Gamma^\lambda_{\mu\sigma} \ .
\label{R}\ee
Under a Weyl transformation (a tool for changing from the Jordan to the Einstein frame), the metric rescales, whereas the Christoffel symbols are independent of the metric and remain unchanged in Palatini gravity. This means that the relation between $K_E^P$ and $K_J^P$  in Palatini formalism is 
\be
K_E^P(\phi )={K_J^P(\phi ) \over \Omega(\phi )}, \qquad 
  K_J^P= K_E^P\, \Omega  \ .
\label{KVEP}
\ee
This is different from the relation between $K_J$ and $K_E$  in the metric formalism shown here in eqs.  \rf{KVE}, \rf{newKVJ}. 

 Now we see two (related) reasons why supergravity in the Jordan frame in eq. \rf{Turin-4}  is incompatible with the Palatini formalism.

1. There is a second term in the equation. \rf{KVE} in $K_E$ which is absent in eq. \rf{KVEP} in the Palatini formalism.

2. In Jordan frame supergravity in eq. \rf{Turin-4}, there is $R((e^a_\mu)_J)$, which indicates that the curvature is given in a form where spin/Christoffel connection is not independent but is a functional of the vierbein/metric shown in eqs. \rf{omegae}, \rf{comp}.

Both points demonstrate that, in the Jordan frame, supergravity has metric-compatible Christoffel symbols that are not independent. We explained in Sec. \ref{Sec:con} that it was necessary, for the consistency of the superconformal theory, to make the spin connection dependent on the vierbein. In the bosonic case, we have a simple relation between the metric and the affine connection with the 
vierbein and spin connection  
\be
g_{\mu\nu}= e_\mu^a \eta_{ab} e_\nu^b\, , \qquad \Gamma^\rho_{\mu \nu }= e^\rho_a (\partial_\mu e^a_\nu+ \omega_\mu{} ^a {}_b \, e^b_\nu) \ .
\ee
Thus, if the spin connection is vierbein dependent, the affine connection is metric and vierbein dependent.
This is why Palatini formalism, which required the affine connection to be independent of the metric, cannot be promoted to a consistent supergravity.

\bibliographystyle{JHEP}
\bibliography{lindekalloshrefs}
\end{document}